\documentclass[english,12pt,sort&compress]{iopart}
\pdfoutput=1 
\usepackage{graphicx}
\usepackage{multirow}

\newcommand{\erf}[1]{\mathrm{Erf}\!\left( {#1} \right)}
\newcommand{\erfc}[1]{\mathrm{Erfc}\! \left( {#1} \right)}
\newcommand{\arcsec}[1]{\mathrm{arcsec}\! \left( {#1} \right)}

\begin{document}
\title{Melting of an Ising Quadrant}
 
\author{P.~L.~Krapivsky}
\address{Department of Physics, Boston University, Boston, MA 02215, USA}
\author{Kirone Mallick}
\address{Institut de Physique Th\'{e}orique CEA, IPhT, F-91191 Gif-sur-Yvette, France}
\author{Tridib Sadhu}
\address{Institut de Physique Th\'{e}orique CEA, IPhT, F-91191 Gif-sur-Yvette, France}

\begin{abstract} 
We consider an Ising ferromagnet endowed with zero-temperature spin-flip dynamics and examine the evolution of the Ising quadrant, namely the spin configuration when the minority phase initially occupies a quadrant while the majority phase occupies three remaining quadrants. The two phases are then always separated by a single interface which generically recedes into the minority phase in a self-similar diffusive manner. The area of the invaded region grows (on average) linearly with time and exhibits non-trivial fluctuations. We map the interface separating the two phases onto the one-dimensional symmetric simple exclusion process and utilize this isomorphism to compute basic cumulants of the area. First, we determine the variance via an exact microscopic analysis (the Bethe ansatz). Then we turn to a continuum treatment by recasting the underlying exclusion process into the framework of the macroscopic fluctuation theory. This provides a systematic way of analyzing the statistics of the invaded area and allows us to determine the asymptotic behaviors of the first four cumulants of the area.
\end{abstract}

\vspace{2pc}

\submitto{\JPA}

\maketitle 

\section{Introduction} 

The studies of fluctuations of growing interface have held a center stage during the
past few decades of the development of non-equilibrium statistical mechanics.
Growing interface appears in numerous physical processes like crystal growth, motion
of grain boundaries under external field, sedimentation, spread of bacterial colony
\textit{etc.} and is a subject of increasing importance from both theoretical and experimental point of view
\cite{Stanley,Halpin1995,Krug1997,Bray94}. Although the microscopic dynamics of the growth
process could be very different, macroscopic fluctuations display a lot of 
universality. This is manifested in continuum descriptions of
the fluctuating interfaces in terms of stochastic partial differential
equations, such as the Edwards-Wilkinson (EW) and Kardar-Parisi-Zhang
(KPZ) equations.

In the last decades, several spectacular advances were made both on the theoretical and
experimental sides \cite{Kriecherbauer2010,Arous2011}, particularly due to
an astounding connection between the ($1+1$) dimensional KPZ equations and random matrix theory
\cite{Baik1999,Johansson2000,Prahofer2000,Prahofer2002,Ferrari2011,Corwin_rev}.
This led to an exact solution of the ($1+1$) KPZ stochastic growth
\cite{Sasamoto2010,Sasamoto2007,Imamura2012,Imamura2013,Amir2011,Calabrese2010,Calabrese2011}.
The universal scaling and the connection to distributions in random matrix theory has been confirmed in a
series of beautiful experiments of kinetic roughening \cite{Takeuchi2010,Takeuchi2011}. 

Thanks to these recent advances the local fluctuations of a growing interface are now well understood in $(1+1)$ dimension. Integral properties are much less explored, however. For instance, one would like to determine the statistics of the area bounded by a growing interface. In experiments, integral characteristics are often more important, and sometimes easier to measure, while from the pure theoretical point of view one might expect that integral characteristics exhibit a  Gaussian statistics even when the local characteristics (like the height of the interface) are non-Gaussian. Even in the situations where the latter is true, one still would like to determine at least the first two moments (the average and the variance) analytically. 

Our purpose here is to analyze the statistics of the simplest integral characteristic of a growing interface---the area under the interface. To this end, we consider an interface growing inside a corner; this geometry has played a prototypical role in previous theoretical works as it allows to represent the interface as an exclusion process on a line \cite{Johansson2000}. Equivalent interpretations of the model are the crystal growth inside a corner, melting of a corner \cite{book}, and the shape of a Young diagram \cite{Vershik1985}.

\begin{figure}
\centering
\includegraphics[scale=0.7]{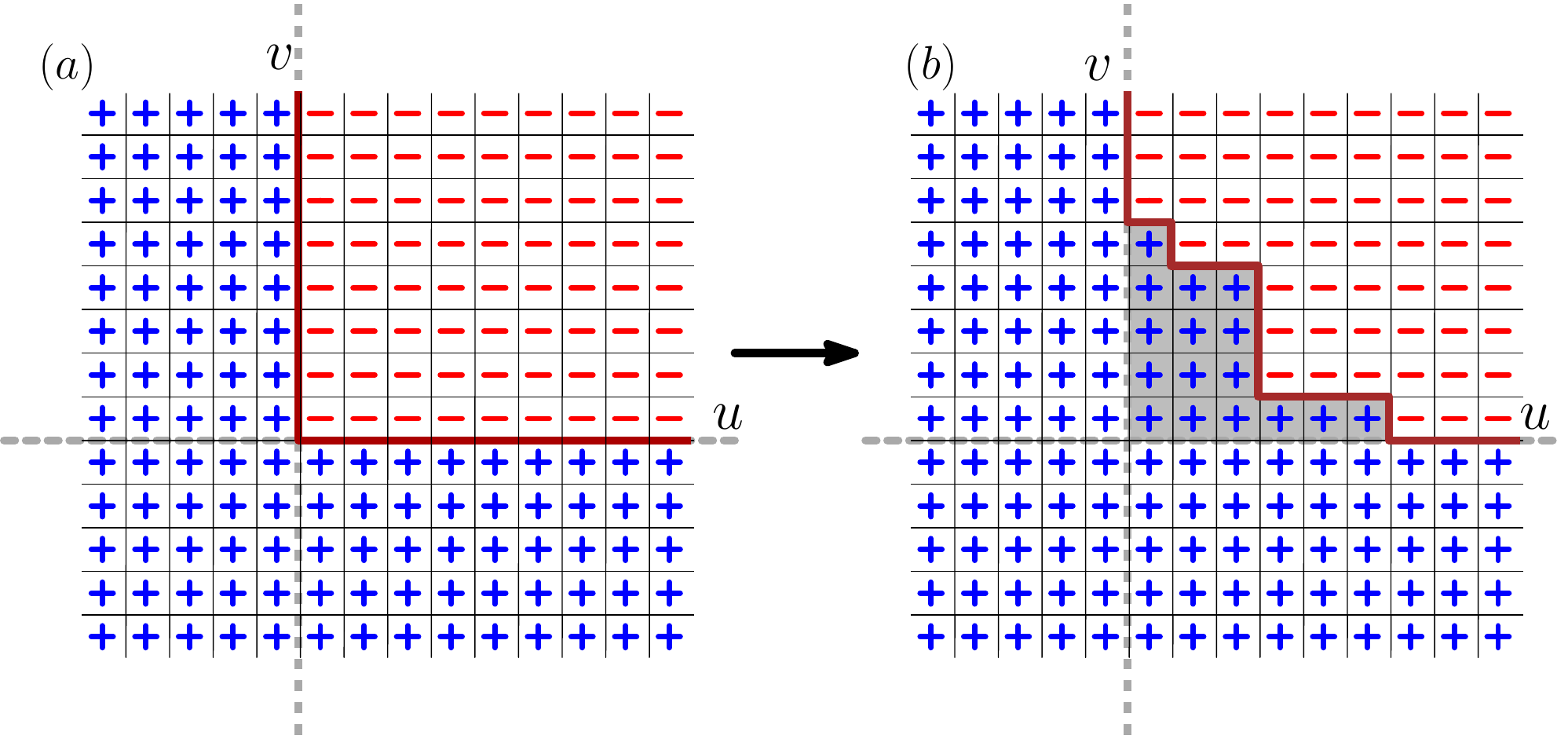}
\caption{ An Ising quadrant defined on a square lattice on the $u$-$v$ plane: $(a)$ the initial
configuration of the Ising spins; $(b)$ a spin configuration at time $T>0$. The interface is the boundary between
the two domains drawn in solid brown line. The shaded area in the second figure denotes the invaded
region where the spins have flipped. In this configuration, the diagonal height $d_T=3\sqrt{2}$ and
the area $A_{T}=16$.} 
\label{fig:def-new}
\end{figure}

More precisely, we analyze the Ising ferromagnet on a square grid endowed with zero-temperature spin-flip
dynamics assuming that initially the minority phase occupies the first quadrant and the majority
phase covers the remaining space (see \Fref{fig:def-new}). The interface separating the phases (initially the surface of
the corner) takes a staircase shape and the area $A_{T}$ of the invaded region  grows linearly with time $T$ on average. The dynamics is stochastic and the shape of the interface varies from one iteration to the other. For instance, the interface may even return to its initial shape (the infinite corner), although the probability of this events quickly decreases with time, viz. it is a stretched exponent in the large time limit [see Eq.~\eref{S_0}]. 

At large times the fluctuations of the interface relative to its size become small and a limiting shape emerges. Limiting shapes of domain boundaries under coarsening dynamics are mostly understood in the frameworks of phenomenological macroscopic description, like the Allen-Cahn equation or the Cahn-Hiliard equation \cite{Bray94,book}. The limiting shapes predicted by these macroscopic descriptions \cite{Julien_LS} differ from the limiting shapes arising in the realm of microscopic descriptions \cite{Alex,Ising_LS,LS_13}.

In the present work we focus on two observables: the height of the interface along the diagonal $d_{T}$ which involves local fluctuations, and the area $A_{T}$ which characterizes global aspects of the fluctuation. Through the mapping onto the one-dimensional symmetric simple exclusion process (SSEP), the height $d_{T}$ corresponds to the integrated current across one bond and $A_{T}$ corresponds to the total displacement of all the particles. We extract the complete statistics of $d_{T}$ from the work of Derrida and Gershenfeld \cite{DG09}. The statistical properties of the area $A_{T}$ are hard to compute because spatial correlations within the entire height profile are required. Our main result is the calculation of the first few cumulants of $A_{T}$. An exact analysis shows that the cumulants exhibit the following long-time asymptotic behavior
\begin{equation}
\label{cumulants}
\langle A_{T}^k \rangle_{c} = C_k T^{(k+1)/2}.
\end{equation}
The computation of the amplitudes $C_k$ becomes involved already for the second cumulant, the variance 
$\langle A_{T}^2 \rangle_{c}= \langle A_{T}^2 \rangle - \langle A_{T} \rangle^2$. We determined the variance of $A_{T}$ using exact microscopic analysis, namely the Bethe Ansatz. To derive the finer statistics of $A_{T}$ we employed a hydrodynamic approach known as the macroscopic fluctuation theory (MFT) which is a powerful general framework for analyzing large deviations in lattice gases \cite{MFT2014,Jona-Lasinio2014,Jona-Lasinio2010,Bodineau2004,Sasorov2014}. Using MFT we additionally calculated the third and fourth cumulants of the invaded area. The expression for the first four cumulants are
\begin{eqnarray}
	\langle A_{T} \rangle_{c} &=& T,  \label{eq:A1_first}\\
  \langle A_{T}^2 \rangle_{c}&=& T^{3/2}\left[ \frac{4}{3}\sqrt{\frac{2}{\pi}}
  \right] \label{eq:A2_first},\\
  \langle A_{T}^3 \rangle_{c}&=& T^{2}\left[ \frac{6\sqrt{3}}{\pi}-2
  \right],  \label{eq:A3_first}\\
\langle A_{T}^{4} \rangle_{c} &=&T^{5/2} ~\frac{32}{5\sqrt{\pi}}
\left[5\sqrt{2} - 4\right.\nonumber\\
  && \qquad \qquad  \left.+\frac{3}{\pi}\left\{ 4-4\sqrt{2}\arccos\left(
    \frac{5}{3\sqrt{3}}\right)-3\sqrt{2}\arccos\left(\frac{1}{3} \right) \right\}\right].
	\label{eq:A4_first}
\end{eqnarray}

We shall present our analysis in the following order.  In \sref{sec:model}, we define the dynamics in detail  and discuss the mapping to the SSEP which will be used to analyze the interface fluctuations. In \sref{sec:fluct} we employ the microscopic analysis which allows us to determine the limiting shape of the interface. Using this analysis we derive
exact expressions for the average and variance of the area. In \sref{sec:mft} we present the formulation of the problem in the framework of the macroscopic fluctuation theory. This allows us to calculate the cumulants of $A_{T}$ up to the
fourth order. In \sref{sec:summary}, we conclude with a brief summary and discuss a few open problems and extensions. Some details of the analysis are relegated to the Appendices. In particular, in \ref{app:fluc hyd} we outline an alternate derivation of the variance $\langle A_{T}^{2} \rangle_{c}$ using fluctuating hydrodynamics and in \ref{app:half-area} we include the analysis of a new observable $H_{T}$, the ``half-area'' of the invaded region, which can be defined for more general initial conditions.

\section{The model and its relation to the exclusion process \label{sec:model}}

We consider an Ising ferromagnet with nearest-neighbor interactions on an infinite square lattice at zero
temperature. In the initial configuration all spins in the first quadrant are
down whereas rest of the spins are up. There is an interface separating the two
oppositely magnetized domains (see \Fref{fig:def-new}). In the starting configuration, the interface is along the positive
coordinate axes, as indicated in the figure. There are two standard spin-flip dynamics for kinetic Ising model---the Glauber and the Metropolis algorithms. At zero temperature, the difference between these two algorithms is small. More precisely, the energy raising flips are forbidden for both algorithms; other flips occur with the same rate according to the Metropolis algorithm, while according to the Glauber algorithm the energy lowering flips proceed twice faster than the energy conserving flips. Staring with our initial configuration, the energy lowering flips never occur and hence the Glauber and the Metropolis algorithms are identical in our setting. In the following, we
set the rate of allowed (energy conserving) flips to unity. 

For the corner initial condition, the plus phase can invade the minus phase, but not the opposite---the three quadrants which are initially occupied by the plus phase cannot be invaded.  The interface separating the plus and minus phases has a staircase shape (see \Fref{fig:def-new}) and it varies from realization to realization. At any moment there is a finite number of `flippable' spins: The total number $N_{-}$ of flippable minus spins, always exceeds by one the total number $N_{+}$ of flippable plus spins. The area $A_T$ (which is nothing but the total number of plus spins in the first quadrant, see \Fref{fig:def-new}) at time $T$ is a random variable; $A_T$ increases by $1$ with rate $N_{-}$ and decreases by $1$ with rate $N_{-}$. Since $N_{-}=1+N_+$, we have $\langle A_T\rangle=T$ at all times. (In contrast, Eqs.~\eref{eq:A2_first}--\eref{eq:A3_first} are valid asymptotically in the $T\rightarrow \infty$ limit.) 

In this paper we are interested in fluctuations of the invaded area. The investigation of the interface dynamics is greatly simplified by the representation in terms of the SSEP. This representation is well known (see \cite{Rost,Liggett,spohn,barma}), so we shall describe it only briefly. 

Before proceeding with the mapping, we set the notation for time: we consider evolution within the time window $[0,T]$ and  we denote an intermediate time by $t$. We also recall that the SSEP is the lattice gas where each site is occupied by at most one particle. In one dimension, each particle hops stochastically with equal unit rate to the neighboring sites on the right and left. Each hopping attempt is successful if the destination site is empty. State of a site $x$ at time $t$ is denoted by a Boolean variable $n_{x}(t)$ which takes value $0$ or $1$ depending on whether the site is empty or occupied. This system of interacting particles has been studied extensively  \cite{Liggett,spohn,Derrida2007,Schutz2001,Chou2011}. 

\begin{figure}
\centering
\includegraphics[scale=0.8]{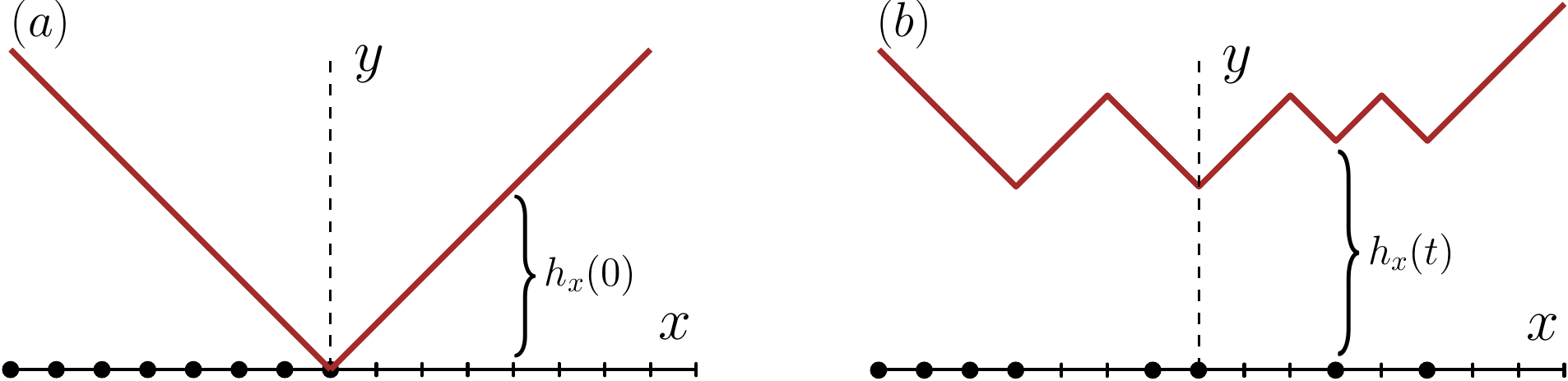}
\caption{ Mapping between the SSEP on a line and an interface. The solid discs on the $x$ axis denote particles and the line denotes the interface corresponding to the particle configuration.
$(a)$ The initial configuration where the particles fill the negative half
chain. The corresponding interface is a right angled wedge. $(b)$ A
configuration at a time $t$ where some particles have
spilled over to the positive side.} 
\label{csp}
\end{figure}

To make the connection to a fluctuating interface we define height  variables $h_x(t)$ which are
related to the occupation variables by
\begin{equation}
	h_{x}(t)-h_{x-1}(t)= 1-2~n_{x}(t).
\label{eq:h tau relation}
\end{equation}
The variable  $h_x(t)$ represents the  height of the interface at position $x$: pictorially this
means that, if a site $x$ is occupied (or empty) then the interface between
$(x-1)$ and $x$ is a straight line going along the co-diagonal (or diagonal)
direction. A schematic of this mapping is shown in \Fref{csp}. In the event of a
particle hopping between two sites, the height at the associated sites changes by $2$. In any
configuration, the height at any two neighboring sites differ by at most $1$. Note that, there
is a unique interface associated with each particle configuration of the exclusion process.

The initial shape of the interface, the corner, corresponds to a step profile in the realm of the exclusion process, namely all sites at $x\le 0$ are occupied, whereas the sites to the right of the origin are empty:
\begin{equation}
	n_{x}(0)=\cases{1 & for $x\le 0$,\\ 0 & for $x>0$.}
	\label{eq:initial SEP}
\end{equation}
The height profile associated to this configuration is $h_{x}(0)=\vert x \vert$. 
One can verify that starting with this configuration, the height at
$x=0$ for any time $t\ge 0$ has a simple expression in terms of the occupation
variables at the right hand side of the origin:
\begin{equation}
	h_{0}(t)=2~\sum_{x=1}^{\infty}n_{x}(t).
	\label{eq:height 0}
\end{equation}

\begin{figure}
\centering
\includegraphics[scale=0.9]{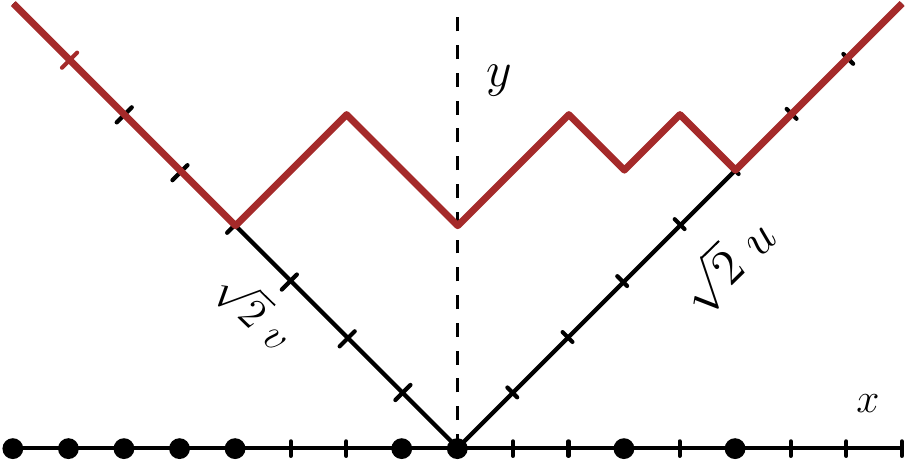}
\caption{An illustration of the rotation of coordinates connecting the interface
  $h_{x}(t)$ on $x$-$y$ plane to the domain boundary of the Ising quadrant on
  the $u$-$v$ plane. The solid brown line denotes the interface. The $\sqrt{2}$ pre-factor is to ensure that $u$ and
$v$ takes integer values.} 
\label{fig:fig3}
\end{figure}

In our original problem of the Ising quadrant, the domain boundary is related to
the interface $\left\{ h_{x}(t) \right\}$ by a rotation of the
coordinates. We use the transformation
\begin{equation}
	u=\frac{x+y}{2} \qquad \textrm{and} \qquad v=\frac{y-x}{2}\,.
	\label{eq:transformation}
\end{equation}
The interface $\{h_{x}(t)\}$ is defined on the $x$--$y$ plane, and the
Ising quadrant is defined on the $u$--$v$ plane. This
corresponds to a $\pi/4$ anti-clockwise rotation and an overall contraction of the
metric by a factor $\sqrt{2}$. A schematic of this transformation is
illustrated in \Fref{fig:fig3}. The contraction in the transformation is to ensure
that each square cell in the Ising model on the $u$--$v$ plane has unit area.

The fluctuations of the domain boundary can be characterized by various
quantities such as  the distance $d_T$ of the domain boundary
from the origin along the diagonal and the change in the area $A_T$ of the
invaded region at time $T$ (see \Fref{fig:def-new}). Using the transformation of the coordinates in
\Fref{csp} it is clear that $d_{T}=h_{0}(T)/\sqrt{2}$ which using Eq.\eref{eq:height 0} yields
\begin{equation*}
	d_T=\sqrt{2}~\sum_{x=1}^{\infty}n_{x}(T).
\end{equation*}
In the exclusion process the sum corresponds to the total current
that has passed through the site at origin up to time $T$. We denote this current by
$Q_{T}$. Then the diagonal height is given by 
\begin{equation}
	d_T=\sqrt{2}~Q_{T}.
	\label{eq:d}
\end{equation}

To define the area $A_T$ in the framework of the exclusion process we note that
\begin{equation}
A_T = \sum_{{\rm all}~{\rm particles}} {\rm displacement},
 \label{Area_displ}
\end{equation} 
\textit{i.e.}, the total displacement of all the particles. Indeed, for the initial condition \eref{eq:initial SEP}, the displacement of the first (right-most) particle is equal to the area of the lowest row of the invaded sites on the $u$-$v$ plane, the displacement of the 2nd particle gives the area of the next row, and so on (see \Fref{fig:def-new} and \Fref{csp}). This proves that the sum on the right-hand side of Eq.~\eref{Area_displ} is really the molten area in the Ising quadrant. It is more convenient,
however, to express the area in terms of the occupation variables $n_x(t)$.
The corresponding expression reads
\begin{equation}
   A_T = \sum_{x = -\infty}^{+\infty}  x~[ n_x(T) - n_x(0)].
 \label{eq:Area}
\end{equation}
Indeed, noting that $A_0=0$ and that any particle hopping to the right (left)
leads to an increase (decrease) of both $A_T$ and the sum in
Eq.\eref{eq:Area} by one,  the formula is established.

\section{Limiting shape and fluctuations \label{sec:fluct}}

The interface generically grows and fluctuates. The relative amplitude of fluctuations compared to the mean
profile of the interface decreases with time, so the re-scaled interface approaches a limiting
shape. To determine the limiting shape we employ a hydrodynamic continuum
description. For the SSEP, the continuum description is the diffusion equation
 \cite{spohn}
\begin{equation}
	\partial_{t} \rho = \partial_{xx}\rho
\label{eq:hydrodynamic}
\end{equation}
describing the evolution of the particle density $\rho(x,t)$. The initial configuration corresponds to a step-like density profile, $\rho(x,0) = \Theta(-x)$, where $\Theta(x)$ is the Heaviside step function.
Solving the diffusion equation with this initial condition, we obtain a solution
in terms of the complementary error function
\begin{equation}
\label{erf}
\rho(x,t)=\frac{1}{2}\,\erfc{\frac{x}{\sqrt{4t}}}.
\end{equation}

Using Eq.\eref{eq:h tau relation} in the continuum limit and the
transformation \eref{eq:transformation}, the mean profile of the interface at time $T$ can be expressed
\cite{Ising_LS} by the following curve on the $(u,v)$ plane,
\begin{equation}
v(u,T)=\int_{u-v}^\infty dz~\frac{1}{2}\,\erfc{\frac{z}{\sqrt{4T}}}.
	\label{eq:main}
\end{equation}
This interface intersects the diagonal line at $u=v=\sqrt{T/\pi}$ and
therefore the average value of the distance $d_{T}$ along the diagonal is
\begin{equation}
\label{C1}
\langle d_{T} \rangle=\frac{\sqrt{2}}{\sqrt{\pi}}\sqrt{T}.
\end{equation}
The angular brackets denote ensemble average.

The average value of the area $A_{T}$ can be deduced from this limiting
shape of the interface. The area under the curve in Eq.\eref{eq:main} at time
$T$ is 
\begin{equation*}
\label{area}
\langle A_T\rangle =\int_0^\infty v(u,T)~du=T 
\end{equation*}
in agreement with Eq.\eref{eq:A1_first}. 

The calculation of the statistics of  $d_{T}$ and $A_{T}$ requires understanding of the fluctuation of the interface around its limiting shape. In Eq.~\eref{eq:d},  the diagonal height $d_{T}$ is essentially the
integrated current $Q_{T}$ through the origin, Eq.~\eref{eq:d}. The statistics of the integrated current has been extensively investigated, see \textit{e.g.} Refs.~\cite{Bodineau2005,Bodineau2006,Bodineau2008,Bunin2012,DG2}. For the step initial condition \eref{eq:initial SEP}, Derrida and Gerschenfeld \cite{DG09} computed the statistics of  $Q_{T}$ using the Bethe Ansatz. Their result leads to the cumulant generating function of $d_{T}$ defined as
$\chi_{T}(\lambda)=\langle \exp[\lambda~d_{T}] \rangle$ (here $\lambda$ is the
fugacity parameter)
\begin{equation}
  \chi_{T}(\lambda)=\frac{\sqrt{T}}{\pi}\int_{0}^{\infty}d\xi \ln\left[
  1+\left( e^{\lambda \sqrt{2}}-1 \right)e^{-\xi^{2}}\right].
	\label{eq:d cum gen}
\end{equation}
The  series expansion of $\chi_{T}(\lambda)$ in powers of $\lambda$
generates all the cumulants of $d_{T}$, which  all  scale  as $\sqrt{T}$. 
The average value \eref{C1} can be retrieved as well.

Compared to $d_{T}$, little is  known about the statistics of  the area  $A_{T}$. In particular, $A_{T}$ contains information of the spatial height-height  correlation of the interface. In terms of the exclusion process,   $A_{T}$  corresponds to the total displacement of all particles. It is simple to verify that
$A_T$ is also the sum of the total current through all the sites on the lattice.
The quantities $A_T$ and $d_T$ are not directly related, for instance for a
fixed $d_{T}$ there is a lower bound on the area, $A_T\geq d_{T}$, but in principle
the area can be arbitrarily large. The only exception is the case of
$d_{T}=0$ when $A_T=0$. This leads to the relation Prob$[A_T = 0]=$
Prob$[d_{T} = 0]$. The latter probability can be extracted from \cite{DG09} to give
\begin{equation}
\label{S_0}
\lim_{T\to\infty} \frac{\ln \textrm{Prob}\left[ A_{T}=0 \right]}{\sqrt{T}} = - \frac{1}{\sqrt{\pi}}\, \zeta
\left(\frac{3}{2}\right) =  -1.473874960...,
\end{equation}
where $\zeta(s)=\sum_{n\geq 1}n^{-s}$ is the zeta function. 

The cumulants of $A_{T}$ are by definition the coefficients of powers of $\lambda$ in the series expansion of the cumulant generating function
\begin{equation}
\mu_{T}(\lambda)=\ln
\langle \exp[\lambda A_T] \rangle.
\label{eq:defmu}
\end{equation}
In other words we have
\begin{equation}
  \mu_T(\lambda)=\lambda \langle A_T \rangle_{c} + \frac{\lambda^2}{2!} \langle
  A_T^2 \rangle_{c}+ \frac{\lambda^3}{3!} \langle A_T^3 \rangle_{c}+\cdots
  \label{eq:mu expansion}
\end{equation}
where, by definition, $\langle A_T^{k}\rangle_{c}$ denotes the $k$th cumulant. 

Using  Eq.~\eref{eq:Area},  all the cumulants of $A_T$ can be
expressed in terms of the equal time correlators of the occupation variables, as
\begin{equation}
	\langle A_T^{k} \rangle_{c}=\sum_{x_{1}}\cdots\sum_{x_{k}} x_{1}\cdots
	x_{k}\langle n_{x_{1}}(T)\cdots n_{x_{k}}(T) \rangle_{c},
\label{eq:AT cumulant}
\end{equation}
for all $k\ge 2$. The $k$-point correlators in the SSEP can be computed using the 
Bethe Ansatz. The resulting exact expressions are difficult to analyze, yet the asymptotic behaviors 
can be extracted using the scaling property \cite{DG09}
\begin{equation}
\label{AT-scaling}
	\langle n_{x_{1}}(T)\cdots n_{x_{k}}(T) \rangle_c
	\simeq T^{(1-k)/2}~G_k\left(\frac{x_{1}}{\sqrt{T}},\cdots,\frac{x_{k}}{\sqrt{T}}\right). 
\end{equation}
Here $G_k(z_{1},\cdots,z_{k})$ is a scaling function. Combining \eref{eq:AT cumulant} and \eref{AT-scaling} yields the general asymptotic time dependence \eref{cumulants}, but the determination of the amplitudes $C_k$ in Eq.~\eref{cumulants} requires a real computation. 

The cumulant generating function which is compatible with \eref{cumulants} must scale as 
\begin{equation}
	\mu_{T}(\lambda)\simeq \sqrt{T}~g\left( \sqrt{T}~ \lambda \right).
	\label{eq:mu scaling}
\end{equation}
The scaling function $g(x)$ does not depend on $T$. \Eref{eq:mu scaling} implies the following large
deviation form
\begin{equation}
 \textrm{Prob}\left( \frac{A_T}{T}=a \right) \asymp \exp\left[-\sqrt{T}~\phi(a)\right].
\label{eq:ldf a}
\end{equation}
The large deviation function $\phi(a)$ is the Legendre transform of 
$\mu_{T}(\lambda)$ \cite{Touchette2009}.

It  might be possible to determine $g(x)$ or $\phi(a)$, but it is a challenging task that has not yet 
been accomplished. Some properties of $\phi(a)$ can be appreciated without an
explicit solution. Near the mean value $a=1$, the function $\phi(a)$ is quadratic,
$\phi(a) \sim (a-1)^{2}$. At large values of $a$, the distribution of $A_{T}$ has a non-Gaussian tail, more precisely 
$\phi(a)\sim a^{3/2}$.  A similar non-Gaussian tail was also found in the distribution of current in the SSEP, see \cite{DG09,Sasorov2014}. To understand the $\phi(a)\sim a^{3/2}$ asymptotic behavior, one can use a heuristic argument \cite{book} which is easier to appreciate using a discrete-time version of the SSEP. In this discrete-time version, particles hop simultaneously at time $t=1,2,3\cdots$ and with equal probabilities $1/2$ to the left and right (whenever the hopping is
possible by the exclusion). The quickest growth of
the total area occurs when all the eligible particles always hop to the right, as
illustrated below:
\begin{eqnarray}
\qquad\qquad\ldots\bullet\bullet\bullet\bullet\bullet\bullet\circ\circ\circ\circ\circ\circ\ldots\nonumber\\
\qquad\qquad\ldots\bullet\bullet\bullet\bullet\bullet\circ\bullet\circ\circ\circ\circ\circ\ldots\nonumber\\
\qquad\qquad\ldots\bullet\bullet\bullet\bullet\circ\bullet\circ\bullet\circ\circ\circ\circ\ldots\nonumber\\
\qquad\qquad\ldots\bullet\bullet\bullet\circ\bullet\circ\bullet\circ\bullet\circ\circ\circ\ldots\nonumber\\
\qquad\qquad\ldots\bullet\bullet\circ\bullet\circ\bullet\circ\bullet\circ\bullet\circ\circ\ldots\nonumber\\
\qquad\qquad\ldots\bullet\circ\bullet\circ\bullet\circ\bullet\circ\bullet\circ\bullet\circ\ldots\nonumber
\end{eqnarray}
This maximal area is readily computed using the representation in Eq.~\eref{Area_displ} to yield
$A_{T}^{\rm max}=1+2+\cdots+T=T(T+1)/2$. The probability of this event is
$2^{-T^2/2}$, leading to
\begin{equation*}
	\ln\left\{\textrm{Prob}\!\left[ A_{T}\simeq \frac{T^2}{2} \right]\right\} \sim -T^{2}.
\end{equation*}
Comparing this with the large-deviation form \eref{eq:ldf a} we see the manifestation of the
non-Gaussian tail of the distribution.

We now use Eq.~\eref{eq:AT cumulant} to  get 
 $\langle A_{T} \rangle$ and to derive an exact expression for the variance
$\langle A_{T}^{2} \rangle_{c}$ at large time $T$. It has been found in \cite{DG09} that at large time $T$,
\begin{eqnarray}
	\langle n_{x}(T) \rangle =\frac{1}{2}\,\erfc{\frac{x}{\sqrt{4T}}},   \label{nx:scaling}\\
	\langle n_{x}(T)n_{y}(T) \rangle_{c} =-\frac{1}{\sqrt{32\pi T}}\exp\left[
	-\frac{(x+y)^{2}}{8 T}
	\right]\,\erfc{\frac{y-x}{\sqrt{8T}}}.
	\label{eq:tt scaling}
\end{eqnarray}
Plugging Eq.\eref{nx:scaling} into $\langle A_T \rangle$ as it is given by 
Eq.\eref{eq:AT cumulant} we arrive at 
\begin{equation}
	\langle A_T \rangle =\int_{-\infty}^{\infty} dx ~
	x~\frac{1}{2}\,\erfc{\frac{x}{\sqrt{4T}}} = T 
\end{equation}
which agrees with the exact result \eref{eq:A1_first}. 

Similarly, from the general expression in Eq.\eref{eq:AT cumulant} the variance of
$A_T$ can be written as
\begin{equation}
\label{variance}
\fl \qquad \langle A_T^2 \rangle_c =  \sum_{x=-\infty}^{\infty} x^2   \langle n_x(T)\rangle
 \left[ 1 -   \langle n_x(T)\rangle \right]
+ 2 \sum_{x=-\infty}^{\infty} ~\sum_{y=x+1}^{\infty} ~x ~y ~\langle n_x(T)  n_y(T)  \rangle_c .
\end{equation}
Combining this result with Eq.\eref{nx:scaling}  and Eq.\eref{eq:tt scaling} we obtain $\langle A_T^{2} \rangle_{c}=C_2T^{3/2}$ with 
\begin{equation*}
\fl \quad C_2 = 2\int_{-\infty}^{\infty}
	d\xi~\xi^2~\erfc{\xi}\erfc{-\xi} -\frac{16\sqrt{2}}{\sqrt{\pi}}
	\int_{-\infty}^{\infty} d\xi \int_{\xi}^{\infty} d\eta
      ~\xi\eta~e^{-(\xi+\eta)^{2}}\erfc{\eta-\xi}
\label{eq:A2 bethe}
\end{equation*}
Evaluating these integrals (with the help of \textit{Mathematica}) we arrive at 
the announced long time asymptotic \eref{eq:A2_first}.

Computing higher cumulants by this technique  gets very cumbersome. The scaling functions for the three or higher point correlators of $n_{x}(T)$ are not known.  In the next section, we use a macroscopic approach to characterize the large fluctuations of the interface at a hydrodynamic scale. This will also enable us 
to calculate the higher order cumulants of $A_T$.

\section{The  macroscopic fluctuation theory approach\label{sec:mft}}

In this section we use the macroscopic fluctuation theory (MFT) developed  by Bertini, De Sole, Gabrielli,
Jona-Lasinio and Landim \cite{MFT2014,Bertini2001,Bertini2002}. This theory provides a general, thermodynamic-like, approach of computing fluctuations and large deviation functions of driven diffusive models (see \cite{MFT2014,Jona-Lasinio2014,Jona-Lasinio2010,Bodineau2004,Sasorov2014} and references therein).

We briefly review the MFT formulation and explain how the MFT can be used  to calculate cumulants of the area
$A_T$ under the fluctuating interface. Then we shall turn to the perturbative analysis. 

\subsection{Application of the MFT to the melting problem}

At a macroscopic scale, the time evolution of the particle density $\rho(x,t)$  in the SSEP is described by a Langevin equation \cite{spohn,Derrida2007,Derrida2011,MFT2014} 
\begin{equation}
  \partial_{t}\rho=\partial_{x}\left[ \partial_{x}\rho
  +\sqrt{\sigma(\rho)}~\eta \right].
  \label{eq:fh}
\end{equation}
Here $\eta(x,t)$ is a Gaussian noise with mean zero and covariance
\begin{equation}
  \langle \eta(x,t)\eta(x^{\prime},t^{\prime})\rangle
  =\delta(x-x^{\prime})~\delta(t-t^{\prime})
  \label{eq:covariance}
\end{equation}
and $\sigma(\rho)$ is the mobility which is 
\begin{equation}
\sigma(\rho)=2\rho(1-\rho)
\label{eq:sigma}
\end{equation}
in the case of the SSEP \cite{DG09,Bodineau2004}. 

The MFT framework \cite{Bertini2001,Tailleur2007,Derrida2007,MFT2014} allows one to assign a probability weight to each history of the density field, evolving according to the Langevin equation \eref{eq:fh}. Furthermore, the MFT provides a scheme of characterizing statistics of any quantity (observable) which is fully determined in terms of the fluctuating density field $\rho(x,t)$. In our case, the area $A_{T}$ can be expressed in terms of the $\rho(x,t)$ by rewriting  Eq.~\eref{eq:Area} in the continuum  limit:
\begin{eqnarray}
  A_T[\rho]&=&\int_{-\infty}^{\infty}dx \left[ \rho(x,T)-\rho(x,0) \right] x.
  \label{eq:AT definition}
\end{eqnarray}
Hence, $A_T[\rho]$ is a functional of the initial and the final density profiles, it does not depend on the profile at intermediate times. Writing the distribution of the final profile $\rho(x,T)$ as a path integral over the density field $\rho(x,t)$ and a conjugate field $\hat{\rho}(x,t)$, we can express (see \ref{app:mft} for details) the generating function of $A_{T}$ as 
\begin{eqnarray}
  \langle e^{\lambda A_T} \rangle= \int
  \mathcal{D}[\rho,\hat{\rho}]~e^{-S\left[ \rho,\hat{\rho} \right]},
  \label{eq:gen fnc}
\end{eqnarray}
with action  
\begin{equation}
S\left[ \rho,\hat{\rho} \right]=-\lambda A_T[\rho]+ \int_{0}^{T}dt\int_{-\infty}^{\infty}dx \left(
    \hat{\rho}~\partial_{t}\rho-H[\rho,\hat{\rho}]\right).
    \label{eq:action}
\end{equation}
Here $H[\rho,\hat{\rho}]$  represents the  Hamiltonian density
\begin{equation}
  H[\rho,\hat{\rho}]=\frac{\sigma(\rho)}{2}\left( \partial_{x}\hat{\rho} \right)^{2}-\left(
  \partial_{x}\rho \right)\left( \partial_{x} \hat{\rho} \right).
\end{equation}
At large $T$, the path integral is dominated by the contribution from the
path $(\rho,\hat{\rho})$ that minimizes the action. Let us denote this optimal
path by $(q,p)$. The associated Euler-Lagrange equations are  
\begin{eqnarray}
  \left( \partial_{t}-\partial_{xx}
  \right)q=-\partial_{x}\left[\sigma(q)\partial_{x} p\right],\label{eq:optimal1}\\
\left( \partial_{t}+\partial_{xx}
\right)p=-\frac{\sigma^{\prime}(q)}{2}\left(\partial_{x} p\right)^2.
\label{eq:optimal}
\end{eqnarray}
The boundary conditions come from the least action condition and by taking into account that the initial density is a step profile. This yields,
\begin{equation}
  p(x,T)=\lambda ~ x \qquad \textrm{and} \qquad q(x,0)=\Theta(-x),
  \label{eq:boundary condition}
\end{equation}
where $\Theta(x)$ is a Heaviside step function.

The saddle point approximation in Eq.\eref{eq:gen fnc} shows that the cumulant generating function $\mu_{T}(\lambda)=\ln\left[ \langle \exp(\lambda A_{T}) \rangle \right]$ is equal to the least action $S[q,p]$. Using Eq.\eref{eq:AT definition} and the optimal equations (\ref{eq:optimal1}--\ref{eq:optimal}), we obtain 
\begin{equation}
 \mu_{T}(\lambda)=\lambda \int_{0}^{T}dt \int_{-\infty}^{\infty}dx\,\, x\,\partial_{t}q
  -\int_{0}^{T}dt \int_{-\infty}^{\infty}dx\,\frac{\sigma(q)}{2}\left( \partial_{x}p \right)^{2}\,.
 \label{eq:mu_0}
\end{equation}
Hence,  the problem of computing the cumulant generating function and the associated large deviation function is equivalent to solving a pair of coupled partial differential equations \eref{eq:optimal1}--\eref{eq:optimal} for two conjugate fields $(q,p)$. The same equations \eref{eq:optimal1}--\eref{eq:optimal} appear in  the analysis of the integrated current \cite{DG2,Krapivsky2012}, in the calculation of large deviation function of density profile in the SSEP \cite{Tailleur2007}, the survival probability of a static target in a lattice gas \cite{MVK14}, etc. The MFT equations have also led to the determination of the statistics of a tagged particle in single-file diffusion \cite{Krapivsky2014}. In all these cases only the boundary conditions are different.

Note that scaling properties can be extracted from  Eq.~\eref{eq:mu_0} without the need of an explicit solution. For example, one can check that $\mu(\lambda)$ is consistent with the scaling in Eq.\eref{eq:mu scaling}, 
confirming  that the $k$th cumulant of $A_{T}$ scales as $T^{(k+1)/2}$ as stated in Eq.\eref{cumulants}.

\subsection{Perturbative analysis}
Exact time-dependent  solutions  of the optimal equations (\ref{eq:optimal1}--\ref{eq:optimal}) have not been found in general;  the tractable settings known so far are those when the optimal solutions are either stationary or traveling waves (see e.g. \cite{Bodineau2004,Bodineau2005,MVK14}). Here, we use a perturbative analysis to solve for few orders in the perturbative expansion in powers of $\lambda$. A similar perturbative analysis was recently applied to the calculation of the variance of integrated current $Q_{T}$ in one-dimensional diffusive systems  \cite{Krapivsky2012}.

The series expansion is around $\lambda=0$. Noting that for $\lambda=0$, the optimal density profile $q(x,t)$ is the solution of the diffusion equation, while the conjugate field $p(x,t)$ vanishes, we seek a perturbative expansion in the form
\begin{eqnarray}
  q=q_{0}+\lambda q_{1}+\lambda^2 q_{2}+\cdots,\\
  p=\lambda p_{1}+\lambda^2 p_{2}+\cdots.
  \label{eq:expansion p q}
\end{eqnarray}
The hydrodynamic solution corresponding to the step initial density profile is
\begin{equation}
  q_{0}(x,t)=\frac{1}{2}\,\erfc{\frac{x}{2\sqrt{t}}}.
  \label{eq:q0}
\end{equation}

The fact that the lowest non-vanishing term in the expansion of $p(x,t)$ is of order $\lambda$, makes it possible to iteratively solve the optimal equations \eref{eq:optimal}. The corresponding equations to each
order has the following general form. The field $p_{k}(x,t)$, at any order $k$, obeys a  time-reversed diffusion equation with a source:
\begin{equation}
\left(\partial_{t}+\partial_{xx}\right)p_{k}=\Gamma_{k},
\end{equation}
where the source term $\Gamma_{k}(x,t)$ depends only on fields of order strictly lower
than $k$. Similarly, the field $q_{k}(x,t)$ satisfies a diffusion equation
\begin{equation}
	\left(\partial_{t}-\partial_{xx}\right)q_{k}=\Delta_{k},
\end{equation}
where the source term $\Delta_{k}(x,t)$ involves $p_k(x,t)$ and fields of order strictly lower than $k$. We seek $q_k$ and $p_k$ in the time window $[0,T]$. The boundary conditions for $q_{k}(x,0)$ and $p_{k}(x,T)$ are determined from 
Eq.~\eref{eq:boundary condition}. A formal solution can be written in terms of the diffusion propagator
\begin{equation}
  g(x,t\vert y, \tau)=\frac{1}{\sqrt{4\pi (t-\tau)}}\,\exp\!\left[
    -\frac{(x-y)^2}{4(t-\tau)}\right],
    \label{eq:g}
\end{equation}
for all $\tau \le t$. We obtain 
\begin{eqnarray}
\fl \qquad	q_{k}(x,t)=\int_{0}^{t}d\tau
	\int_{-\infty}^{\infty}dy~\Delta_{k}(y,\tau)~g(x,t \vert
	y,\tau)+\int_{-\infty}^{\infty}dy~q_{k}(y,0)~g(x,t \vert y,
	0)\label{eq:formal q}\\
\fl \qquad p_{k}(x,t)=-\int_{t}^{T}d\tau
	\int_{-\infty}^{\infty}dy~\Gamma_{k}(y,\tau)~g(y,\tau \vert
	x,t)+\int_{-\infty}^{\infty}dy~p_{k}(y,T)~g(y,T \vert x, t).
\label{eq:formal p}
\end{eqnarray}

Our goal is to use this perturbative solution to compute the series expansion  of $\mu(\lambda)$ and therefore the cumulants of $A_{T}$, see Eq.~\eref{eq:mu expansion}. The $k$th cumulant
$\langle A_{T}^{k} \rangle_{c}$  will depend only on the solution of $p(x,t)$ and $q(x,t)$ up to the $(k-1)$th order at most; {\it e.g.}, the average $\langle A_T\rangle$ is solely determined in terms of $q_{0}(x,t)$.

Using Eq.\eref{eq:optimal1} we replace $\partial_{t}q$ by $\partial_x[\partial_x q -\sigma(q)\partial_x p]$ in the first integral \eref{eq:mu_0}. Integrating by parts we recast Eq.\eref{eq:mu_0} into 
\begin{equation}
\mu_T(\lambda)=\lambda T+\int_{0}^{T}dt \int_{-\infty}^{\infty}dx\,
  \frac{\sigma(q)}{2}\left( \partial_{x}p \right)\left[ 2\lambda
  -\partial_{x}p \right]. 
\label{eq:mu 2}
\end{equation}
Because  the lowest non-vanishing term in the expansion of $p(x,t)$ is
of order $\lambda$, the second integral is of order $\lambda^2$ or higher. Thus $\langle A_T\rangle = T$ and 
substituting the perturbative expansion \eref{eq:expansion p q} into \eref{eq:mu 2} we deduce the following cumulants
\begin{eqnarray}
  \frac{1}{2!}\langle A_{T}^{2} \rangle_{c} &=& \frac{1}{2}\int_{0}^{T}dt
  \int_{-\infty}^{\infty}dx ~\sigma_{0}, \label{eq:A2}\\
  \frac{1}{3!}\langle A_{T}^{3} \rangle_{c} &=& \frac{1}{2} \int_{0}^{T}dt
  \int_{-\infty}^{\infty}dx ~\sigma_{1}, \label{eq:A3}\\
  \frac{1}{4!}\langle A_{T}^{4} \rangle_{c} &=& \frac{1}{2} \int_{0}^{T}dt
  \int_{-\infty}^{\infty}dx
  \left[\sigma_{2}-\sigma_{0}(\partial_{x}p_{2})^{2}\right].
  \label{eq:A4}
\end{eqnarray}
Here we used that $\partial_x p_1=1$ which is proved later in Eq.\eref{eq:p1}.
Here $\sigma_{k}\equiv \sigma_{k}(x,t)$ is the $k$th order term in the expansion of
$\sigma[q(x,t)]$ in powers of $\lambda$. From $\sigma(q)=2q(1-q)$ one finds 
\begin{eqnarray}
\label{eq:sigma012}
  \sigma_0   &=  2q_0\left[ 1-q_0\right],  \nonumber\\ 
  \sigma_{1} &=  2 q_{1}\left[1-2q_0\right], \\
  \sigma_{2} &=  2 q_{2}\left[ 1- 2q_0\right]-2 q_{1}^2. \nonumber
\end{eqnarray}

We now outline the computation of the second, third, and the fourth cumulants.  

\subsubsection{Variance of the Area:}
Since $\sigma_0=2q_0(1-q_0)$, see Eq.\eref{eq:sigma012}, it is clear that
the second cumulant $\langle A_T^{2} \rangle_{c}$ in Eq.\eref{eq:A2} involves only the zeroth order
solution $q_{0}(x,t)$ which is given by 
Eq.\eref{eq:q0}. Using the rescaled variable $\xi=x/2\sqrt{T}$, we obtain 
\begin{equation}
\label{A2:MFT}
	\langle A_T^2 \rangle_{c}= T^{3/2}\,
  \frac{2}{3} \int_{-\infty}^{\infty}d\xi~ \erfc{\xi}\erfc{-\xi}=T^{3/2}\left[ \frac{4}{3}\sqrt{\frac{2}{\pi}} \right]. 
\end{equation}
Numerically evaluating the expression yields $\langle A_T^2 \rangle_{c}\simeq 1.063~T^{3/2}$.
Interestingly, the integral expression in Eq.\eref{A2:MFT} is much simpler
than the integral in Eq.\eref{eq:A2 bethe} obtained from the Bethe Ansatz. The answers are, of course, the same, so we have another derivation of the asymptotic Eq.\eref{eq:A2_first} for the variance. The same result can be deduced by linearizing Eq.\eref{eq:fh} and  assuming small fluctuations around the hydrodynamic solution
(see \ref{app:fluc hyd} for details).

\subsubsection{Skewness of the Area (Third cumulant):}

The expression for the third cumulant requires the knowledge of  $q_{1}(x,t)$ and $p_{1}(x,t)$. Substituting the perturbative expansion into Eqs.~(\ref{eq:optimal1}--\ref{eq:optimal}) yields 	
\begin{equation*}
 \left(\partial_{t}-\partial_{xx}\right)q_{1}=-\partial_{x}\left[\sigma_{0}\partial_{x}p_1\right], \quad 
 \left(\partial_{t}+\partial_{xx}\right)p_{1}=0,
\end{equation*}
at the first order in $\lambda$. The boundary conditions (\ref{eq:boundary condition}) become 
\begin{equation}
\label{BC:1}
  p_{1}(x,T)=x \qquad \textrm{and} \qquad q_{1}(x,0)=0 \, . 
\end{equation}
Solving $\left(\partial_{t}+\partial_{xx}\right)p_{1}=0$ subject to the first boundary condition in Eq.~\eref{BC:1} we get
\begin{equation}
p_{1}(x,t)=x \quad\textrm{for}\quad  0\le t\le T. 
\label{eq:p1}
\end{equation}
Using this result we simplify the governing equation for $q_1(x,t)$ to 
\begin{equation}
  \label{eq:q1 eq}
\left(\partial_{t}-\partial_{xx}\right)q_{1}=-\partial_{x}\sigma_{0}
\end{equation}
whose solution reads
\begin{equation}
  q_{1}(x,t)=2\int_{0}^{t}d\tau\int_{-\infty}^{\infty}dy ~g(x,t\vert
  y,\tau)~\erf{\frac{y}{2\sqrt{\tau}}}g(y,\tau\vert 0,0),
  \label{eq:q1 final}
\end{equation}
using $\partial_{y}\sigma_0(y,\tau)=-2~\erf{y/2\sqrt{\tau}}g(y,\tau\vert 0,0)$.

Using Eq.\eref{eq:q1 final} one can determine the third cumulant  \eref{eq:A3}. The computations are a bit lengthy (see \ref{app:A3}), but the final result is neat
\begin{equation}
  \langle A_{T}^{3}\rangle_c = T^{2} \left[
  \frac{6\sqrt{3}}{\pi} -2\right]\simeq 1.308~T^2.
  \label{eq:A3 second}
\end{equation}
The non-vanishing of the skewness indicates that $A_{T}$ is a non-Gaussian
random variable, whereas the
Edwards-Wilkinson equation predicts a Gaussian behavior (see \ref{app:fluc hyd} for more details). 

\subsubsection{Flatness of the Area (Fourth cumulant):}
The fourth cumulant requires solutions for $p(x,t)$ and $q(x,t)$ up to the second order in $\lambda$. Recalling that $\partial_{x}p_{1}=1$ [see Eq.~\eref{eq:p1}], we write these equations as  
\begin{eqnarray}
\left(\partial_{t}-\partial_{xx}\right) q_{2} &=& -\partial_x\left[
	\sigma_{1}+\sigma_{0}\partial_{x}p_{2} \right].
	\label{eq:q2 equation} \\
\left(\partial_{t}+\partial_{xx}\right)p_2 &=& 2q_{0}-1,
\label{eq:p2 equation}
\end{eqnarray}
The boundary conditions read
\begin{equation}
	p_{2}(x,T)=0 \qquad \textrm{and} \qquad q_{2}(x,0)=0.
	\label{eq:p2 q2 boundary}
\end{equation}
Combining the formal solution (\ref{eq:formal p}) and the expression \eref{eq:q0}  for  $q_{0}(x,t)$ we get
\begin{equation}
	p_{2}(x,t)=\int_{t}^{T}d\tau\int_{-\infty}^{\infty}dy
	~\erf{\frac{y}{2\sqrt{\tau}}}~g(y,\tau \vert x, t).
	\label{eq:p2}
\end{equation}
Similarly, the solution of Eq.\eref{eq:q2 equation} is given by 
\begin{eqnarray}
\label{eq:q2}
\fl~	q_{2}(x,t)\!=\!\int_{0}^{t}d\tau \int_{-\infty}^{\infty}dy
~\left\{2 q_{1}(y,\tau)[1-2q_{0}(y,\tau)]+\sigma_{0}(y,\tau)\partial_{y}p_{2}(y,\tau)\right\}\partial_{y}g(x,t\vert y,\tau).
\end{eqnarray}
In deriving Eq.\eref{eq:q2} we have used the expressions for $\sigma_1(x,t)$ from Eqs.~\eref{eq:sigma012} and the identity  $\partial_{x}g(x, t \vert y, \tau)=-\partial_{y}g(x, t \vert y, \tau)$.

Using $p_2(x,t)$ and $q_2(x,t)$, Eqs.~\eref{eq:p2}--\eref{eq:q2}, and the previously derived $q_0, q_1, p_1$ we can determine the forth cumulant.
The computation of the integrals is quite involved, so the 
details are deferred to \ref{app:A4}. Here we just state the final result: the fourth cumulant has a closed form expression
\begin{eqnarray}
\fl	\langle A_{T}^{4} \rangle_{c} =T^{5/2} \frac{32}{5\sqrt{\pi}}\left[5\sqrt{2}-4
  +\frac{3}{\pi}\left\{ 4-4\sqrt{2}\arccos\left(
    \frac{5}{3\sqrt{3}}\right)-3\sqrt{2}\arccos\left(\frac{1}{3} \right) \right\}\right].
    \label{eq:A4 final}
\end{eqnarray}
Numerically evaluating the expression yields $\langle A_{T}^{4} \rangle_{c} \simeq 1.497~T^{5/2} $.

The perturbative expansion could be pushed forward to calculate higher order cumulants. The analysis gets more and more cumbersome and a systematic scheme is required. The fourth cumulant $\langle A_T^4 \rangle_c$ involves six layers of complicated integrals (see \ref{app:A4}), so completing the task and establishing \eref{eq:A4 final} was rather unexpected. This suggests that the problem may have some integrable structure that would make it fully solvable. Besides, an intricate recursive structure in the solutions
for $p_{k}(x,t)$ and $q_{k}(x,t)$  in terms of graphs emerges
\cite{sylvain}. This deserves to be explored further---the hope is to find a pattern
in the expression for the cumulants which may help in estimating the expression
for the entire cumulant generating function.  Nevertheless, the results of 
this section have shown that the MFT is a powerful computational tool to explore the
 statistical properties of an observable that can be written as a functional of
the solution of  a non-linear, fluctuating, hydrodynamic equation. 

\section{Discussion \label{sec:summary}}

We considered an Ising ferromagnet endowed with zero-temperature spin-flip dynamics. The Ising quadrant melts, and we studied the statistics of the total melted area $A_T$.  The total area is a global observable of the melted region that involves the multiple-point correlations of the interface height. We focused on a symmetric dynamics in which deposition and evaporation  events occur with the same rates. The local behavior of the height of the interface can be described by the Edwards-Wilkinson growth model, a linear and tractable stochastic equation. However, the statistics of  $A_T$  requires the knowledge of  the spatial fluctuations and correlations of the interface. We calculated the average, the variance, the skewness and the flatness of $A_T$ by solving perturbatively the optimal equations of the MFT. We also used exact microscopic calculations based on the Bethe Ansatz to determine the average and the variance and found the same results. The MFT provides a systematic computational scheme that can be carried over to higher orders. Besides, the calculations based on MFT are already simpler at the second order ({\it i.e.},  for the variance), compared to the Bethe Ansatz.

Our initial goal was to establish a closed expression for the cumulant generating function $\mu_T(\lambda)$ of  $A_T$. We have only derived the Maclaurin  expansion of that function up to the fourth order; by a scaling argument, we also know its leading behavior in the $\lambda \to \infty$ limit. The expressions \eref{eq:A1_first}--\eref{eq:A4_first} for the cumulants up to the fourth order do not seem to suggest a conjectural form of the higher cumulants. We leave this problem for future investigations. In particular, the presence of a recursive structure in the perturbative analysis  of the MFT equations \cite{sylvain} hints at some integrability property that leaves the hope that these equations could be solvable. 

The total melted area is the most basic global observable characterizing the melted region. There are other global observables, e.g., the total number of flippable plus spins $N_+$ and the total number of flippable minus spins $N_-$. It suffices to consider $N_+$ as $N_-=1+N_+$. The statistics of $N_+(T)$ has not been probed. The average growth is not difficult to deduce \cite{LS_13}, 
\begin{equation}
\langle N_+\rangle = \sqrt{\frac{2T}{\pi}}\,.
\end{equation}
The variance is unknown, although one anticipates that $\langle N_+^2\rangle_c = B_2\sqrt{T}$, and generally $\langle N_+^k\rangle_c = B_k\sqrt{T}$. 

One can also modify the underlying Ising model. For instance, instead of the Ising ferromagnet with nearest-neighbor interactions, one can consider more general ferromagnets, e.g., with next-nearest-neighbor (still ferromagnetic) interactions. The mapping of the interface problem onto a one-dimensional diffusive lattice gas still holds \cite{LS_13}, but the corresponding lattice gas has a density-dependent diffusion coefficient and a rather complicated mobility \cite{RP_12}. The limiting shape and hence the average area are known \cite{LS_13}, while even the computation of the variance appears very difficult. 

Perhaps a more fundamental change is to study the same Ising ferromagnet with nearest-neighbor interactions, but in the presence of a magnetic field favoring the majority phase. The corresponding particle system is the totally asymmetric simple exclusion process (TASEP).  A huge corpus of theoretical results has been derived for the TASEP (equivalently, for the KPZ interface in 1+1 dimensions), there are also experimental realizations (see e.g. \cite{Halpin1995,Kriecherbauer2010,Corwin_rev,Takeuchi2011} and  references therein). To the best of our knowledge, the observable corresponding to the total area  has not been studied. The average area $\langle A_T \rangle = T^2/6$ is well-known \cite{Rost}, and the growth of the variance $\langle A_T^2 \rangle_c \sim  T^{7/3}$ can be estimated using a scaling argument \cite{sylvain,Olejarz2013}. The precise calculation of the variance (and of higher moments) is an open problem. We recall  that the MFT scheme can not be applied {\it per se} to the TASEP, which  is a non-diffusive system.

The extension to higher dimensions is an outstanding challenge. Scaling laws for the average volume and its variance can be expressed in terms of scaling exponents of the continuous growth models (see \ref{app:fluc hyd} and \cite{Olejarz2013}). Little is known about the scaling exponents above $1+1$ dimensions \cite{Halpin2012,Halpin2014}, however, especially in the situation with a magnetic field (when the growth process is in the KPZ universality class). The absence of a mapping of the interface dynamics onto a simple lattice gas is another barrier which currently prevents us from applying the MFT to the computation of the statistics of the growing volume.

\section*{Acknowledgments}

We are grateful to S.~Prolhac for discussions. We thank S. Mallick for a critical reading of the manuscript. The research of PLK was supported by a grant from BSF.

\appendix

\section{Derivation of the MFT action and the associated Euler-Lagrange equation \label{app:mft}}

At a macroscopic scale the time evolution of the coarse-grained density profile is governed by the fluctuating hydrodynamic equation \eref{eq:fh}. Considering all possible evolution of $\rho(x,t)$ in the time interval $[0,T]$, the moment generating function of the area $A_{T}$ can be written as a path integral
\begin{equation}
 \langle e^{\lambda A_T} \rangle =\int \mathcal{D}[\rho]\, e^{\lambda A_T[\rho]}\, \Big\langle \delta\Big( \partial_t \rho -\partial_x \Big[\partial_x\rho+\sqrt{\sigma(\rho)}\eta\Big] \Big) \Big\rangle.
\end{equation}
The Dirac delta function $\delta(z)$ is to ensure that contributions only come from the paths that follow the Eq.\eref{eq:fh}. The average $\langle \rangle$ is over the history of the noise $\eta(x,t)$. 

The delta function can be replaced by a
path integral over a conjugate field $\hat{\rho}(x,t)$, which leads to
\begin{equation}
\fl \qquad  \langle e^{\lambda A_T} \rangle =\int \mathcal{D}[\rho,\hat{\rho}]\, e^{\lambda A_T[\rho]}\, 
\Big\langle e^{-\int_0^T dt \int_{-\infty}^{\infty}dx \hat{\rho}\left[\partial_t \rho -\partial_x \left(\partial_x\rho+\sqrt{\sigma(\rho)}~\eta \right) \right]} \Big\rangle.
\end{equation}
Using integration by parts and assuming that $\partial_x\rho$ and $\sigma(\rho)$ vanish at $x\rightarrow \pm \infty$, the expression yields
\begin{equation}
\fl \qquad \langle e^{\lambda A_T} \rangle =\int \mathcal{D}[\rho,\hat{\rho}]\, e^{\lambda A_T[\rho]-\int_0^T dt \int_{-\infty}^{\infty}dx \left[ \hat{\rho} \partial_t\rho +(\partial_x \rho)(\partial_x\hat{\rho}) \right]}\, \Big\langle 
 e^{-\int_0^T dt\int_{-\infty}^{\infty}dx\sqrt{\sigma(\rho)}(\partial_x\hat{\rho})\eta} \Big\rangle.
 \label{eq:B3}
\end{equation}
Since $\eta(x,t)$ is a Gaussian noise with covariance \eref{eq:covariance}, the average is 
\begin{equation*}
	\Big\langle e^{-\int_0^T dt\int_{-\infty}^{\infty}dx\sqrt{\sigma(\rho)}(\partial_x\hat{\rho})\eta} \Big\rangle= 
	e^{\int_0^T dt\int_{-\infty}^{\infty}dx \frac{\sigma(\rho)}{2}(\partial_x\hat{\rho})^2}.
\end{equation*}
Substituting this in Eq.\eref{eq:B3} we obtain the announced result, Eq.~\eref{eq:action}, for the action.

At large $T$, this effective action grows as $\sqrt{T}$~ \cite{DG2} and the path integral is dominated by its saddle point.  We now minimize the action. Denote by $(q,p)\equiv (\rho,\hat{\rho})$ the path that minimizes the action and take a small variation $(\delta\rho,\delta \hat{\rho})$ around this path. The change in action $S[\rho,\hat{\rho}]$ in Eq.~\eref{eq:action} corresponding to this variation is
\begin{eqnarray}
	\delta S=\int_{-\infty}^{\infty}dx \left\{ -\lambda
	\frac{\delta A_{T}[q]}{\delta q(x,0)} -p(x,0) \right\}\delta \rho(x,0)\nonumber\\
	\qquad+\int_{-\infty}^{\infty}dx \left\{ -\lambda
	\frac{\delta A_{T}[q]}{\delta q(x,T)}+p(x,T) \right\}\delta
	\rho(x,T)\nonumber\\
	\qquad+\int_{0}^{T}dt \int_{-\infty}^{\infty}dx \left\{
	-\partial_{t}p-\frac{\sigma^{\prime}(q)}{2}\left( \partial_{x}p
	\right)^{2} -\partial_{xx}p\right\}\delta \rho(x,t)\nonumber\\
	\qquad+\int_{0}^{T}dt \int_{-\infty}^{\infty}dx \left\{
	\partial_{t}q+\partial_{x}\left[\sigma(q) \partial_{x}p - \partial_{x}q
	\right]\right\}\delta
	\hat{\rho}(x,t),
	\label{eq:action variation}
\end{eqnarray}
where the functional derivatives are taken at the optimal path $(q,p)$.

For the action $S[q,p]$ to be minimum, the variation must vanish. Since $\delta \rho(x,t)$ and $\delta\hat{\rho}(x,t)$ are arbitrary, the terms inside the curly brackets in the last two integrals in Eq.~\eref{eq:action variation} must vanish. This leads to the governing equations \eref{eq:optimal1}--\eref{eq:optimal}.

The first integral in Eq.\eref{eq:action variation} vanishes due to the fixed initial profile  $\rho(x,0)=\Theta(-x)$, which implies that the variation $\delta\rho(x,0)=0$. This also provides the first boundary condition in Eq.\eref{eq:boundary condition}. Vanishing of the second integral leads to the second boundary condition. As the final profile $\rho(x,T)$ is fluctuating, the variation $\delta \rho(x,T)$ is arbitrary. Thus the term inside the curly brackets in the second integral in Eq.\eref{eq:action variation} must vanish, leading to the condition
\begin{equation*}
	p(x,T)=\lambda\frac{\delta A_{T}[q]}{\delta q(x,T)}.\label{eq:pT}
\end{equation*}
Using the expression for $A_T[q]$ from Eq.\eref{eq:AT definition} we obtain the boundary conditions \eref{eq:boundary condition}.

\section{Derivation of the third cumulant of the area $A_{T}$ \label{app:A3}}

To compute the third cumulant \eref{eq:A3} we need to know the expansion of 
$q(x,t)$ up to the first order. Substituting $q_{0}$ from Eq.\eref{eq:q0} and 
$q_{1}$ from Eq.\eref{eq:q1 final} into Eq.\eref{eq:A3}  we obtain
\begin{equation*}
\fl \frac{1}{3!}\langle A_{T}^{3} \rangle_c = 2\int_{0}^{T}dt
 \int_{-\infty}^{\infty}dx\int_{0}^{t}dt^{\prime}\int_{-\infty}^{\infty}dy~\erf{
   \frac{x}{2\sqrt{t}}}
 g(x,t\vert y,
  t^{\prime})~\erf{\frac{y}{2\sqrt{t^{\prime}}}}g(y,t^{\prime} \vert 0,0).
\end{equation*}
To simplify the integral on the right-hand side we use new variables
$z=y/2\sqrt{t^{\prime}}$, $\omega=(x-y)/2\sqrt{t-t^{\prime}}$, and
$\alpha=t^{\prime}/t$ and after straightforward manipulations we get
\begin{equation}
  \label{eq:A3 integral}
\frac{1}{3!}\langle A_{T}^{3}\rangle_c=T^{2}\,\frac{1}{\pi}\int_0^1 d\alpha\,I(\alpha)
\end{equation}
where we have used the shorthand notation
\begin{equation}
	I(\alpha)=\int_{-\infty}^{\infty}dz\int_{-\infty}^{\infty}d\omega\exp\left[
	-\omega^{2}-z^2	\right]\erf{z}\erf{\omega\sqrt{1-\alpha}+z\sqrt{\alpha}}.
\label{Ia}
\end{equation}
It turns out that 
\begin{equation}
\label{Ia_sin}
I(\alpha)=2\arcsin\left[\frac{\sqrt{\alpha}}{2} \right].
\end{equation}
To establish this identity we first note that for $\alpha=0$ both sides in Eq.~\eref{Ia_sin} vanish. Next, we differentiate both sides with respect to $\alpha$ and show that the outcomes are identical. The derivative of the left-hand side of Eq.~\eref{Ia_sin} gives a Gaussian integral convoluted with an error function,
\begin{eqnarray}
\frac{d I(\alpha)}{d\alpha}=\frac{2}{\sqrt{\pi}}\int_{-\infty}^{\infty}dz\int_{-\infty}^{\infty}d\omega\,
\exp\left[-z^2-\omega^2-\left(
	z\sqrt{\alpha}+\omega\sqrt{1-\alpha}
	\right)^2\right]\nonumber\\ \qquad \qquad\qquad\qquad\qquad\qquad\left(
	\frac{z}{2\sqrt{\alpha}}-\frac{\omega}{2\sqrt{1-\alpha}}
	\right)\erf{z}. 
\end{eqnarray}
The integral over $\omega$  is Gaussian integral and we compute it first. The integrals over $z$ are then computed through integration by part. The result is
\begin{equation*}
\frac{d I(\alpha)}{d\alpha}=\frac{1}{\sqrt{\alpha(4-\alpha)}}.
\end{equation*}
Taking derivative of the expression on the right-hand side in Eq.~\eref{Ia_sin} with respect to
$\alpha$ yields the same result. This completes the proof of the identity \eref{Ia_sin}.

Using Eq.\eref{Ia_sin} we recast Eq.\eref{eq:A3 integral} into
\begin{equation}
  \frac{1}{3!}\langle A_{T}^{3}\rangle_c=T^{2} \left[\frac{2}{\pi}\int_{0}^{1}d\alpha
  ~\arcsin\left( \frac{\sqrt{\alpha}}{2} \right)\right].
\end{equation}
Computing the integral leads to the result reported in Eq.\eref{eq:A3 second}.

\section{Derivation of the fourth cumulant of the area $A_{T}$ \label{app:A4}}

The expression for the fourth cumulant \eref{eq:A4} involves
$\partial_{x}p_{2}$ and $q_0, q_1, q_2$. First we use Eq.\eref{eq:p2} to calculate
\begin{equation}
	\partial_{x}p_{2}(x,t)=\int_{t}^{T}d\tau\int_{-\infty}^{\infty}dy
	~\erf{\frac{y}{2\sqrt{\tau}}}~\partial_{x}g(y,\tau \vert x, t).
	\label{eq:dxp2}
\end{equation}
Utilizing the identity
$\partial_{x}g(y,\tau \vert x, t)=-\partial_{y}g(y,\tau \vert x, t)$ and the integration by parts to transfer the partial derivative on the $\erf{y/2\sqrt{\tau}}$ we arrive at 
\begin{equation}
	\partial_{x}p_{2}(x,t)=2\int_{t}^{T}d\tau \int_{-\infty}^{\infty}dy
	~g(y,\tau\vert 0,0)~g(y,\tau\vert x, t),
\end{equation}
where we additionally used the identity $\partial_{y}\erf{y/2\sqrt{\tau}}=2 g(y,\tau \vert 0,0)$. The integral over $y$ is a Gaussian integral which we compute and get 
\begin{equation}
	\partial_{x}p_{2}(x,t)=2\int_{t}^{T}d\tau ~g(0,2\tau \vert x, t).
	\label{eq:partial p2}
\end{equation}
This last integral can be evaluated, but it proves more convenient to keep the integral
form.

We will also need an alternative formula for $q_1(x,t)$. In Eq.~\eref{eq:q1 final} the integration over $y$ can be performed using the following general identity \cite{Prudnikov}
\begin{equation}
	\int_{-\infty}^{\infty}dz~\erfc{\frac{z}{\beta}}
	\exp[-\alpha(z-x)^2]=\sqrt{\frac{\pi}{\alpha}}~\erfc{x\sqrt{\frac{\alpha}{1+\alpha~\beta^2}}}	,
\end{equation}
which leads to
\begin{equation}
	q_1(x,t)=2~t~g(x,t\vert 0,0) \int_0^1 dr ~\erf{\frac{x\sqrt{r}}{2\sqrt{t(2-r)}}}.
	\label{eq:q1 new}
\end{equation}

We now use $\sigma_2$ from Eq.~\eref{eq:sigma012} and re-write the expression \eref{eq:A4} for the fourth cumulant as a sum of three terms
\begin{eqnarray}
	\frac{1}{4!} \langle A_{T}^{4} \rangle_{c} =I_{1}-I_{2}-I_{3},
	\label{eq:A4 sum}\\
	I_{1}=\int_{0}^{T}dt \int_{-\infty}^{\infty}dx\,q_{2}\left[ 1
	-2q_{0} \right]\label{eq:I1}\\
	I_{2}=\int_{0}^{T}dt \int_{-\infty}^{\infty}dx\,q_{1}^{2}
	\label{eq:I2}\\
	I_{3}=\frac{1}{2}\int_{0}^{T}dt
	\int_{-\infty}^{\infty}dx~\sigma_{0}\left[\partial_{x}p_{2}\right]^{2}
	\label{eq:I3}
\end{eqnarray}

\subsection*{Computation of $I_1$}
Using Eq.\eref{eq:q2} we re-write $I_1$ as
\begin{eqnarray}
\fl	I_{1}=2\int_{0}^{T}dt \int_{-\infty}^{\infty}dx\int_0^t d\tau \int_{-\infty}^{\infty}dy\,q_1(y,\tau)\left[1-2q_0(y,\tau)\right]\left[1-2q_{0}(x,t) \right]\partial_y g(x,t \vert y,\tau)\nonumber\\
\fl\qquad	+\int_{0}^{T}dt \int_{-\infty}^{\infty}dx\int_0^t d\tau \int_{-\infty}^{\infty}dy\, \sigma_0(y,\tau)\partial_y p_2(y,\tau)\left[ 1
	-2q_{0}(x,t) \right]\partial_y g(x,t \vert y,\tau).
	\label{eq:C8}
\end{eqnarray}
It turns out that the second integral in Eq.\eref{eq:C8} is equal to $2I_3$. To show this we use
\begin{equation*}
\int_0^T dt \int_0^t d\tau\equiv \int_0^T d\tau \int_\tau^T dt
\end{equation*}
and re-write the second integral in Eq.\eref{eq:C8}  as
\begin{equation*}
\fl	\qquad \int_0^T d\tau \int_{-\infty}^{\infty}dy ~\sigma_0(y,\tau) \partial_y p_2(y,\tau) \left\{\int_\tau^T dt \int_{-\infty}^{\infty}dx \left[1- 2q_0(x,t) \right]\partial_y g(x,t \vert y,\tau) \right\}.
\end{equation*}
The term inside the curly brackets is equal to $\partial_y p_2(y,\tau)$, see Eq.\eref{eq:dxp2}. This shows that the second integral in \eref{eq:C8} is indeed equal to $2 I_3$.

Thus we can re-arrange Eq.~\eref{eq:C8} as
\begin{equation*}
\fl	I_{1}-2I_{3}=
2\int_{0}^{T}dt \int_{-\infty}^{\infty}dx\int_0^t d\tau \int_{-\infty}^{\infty}dy\, q_1(y,\tau)\left[1-2q_0(y,\tau)\right]\left[ 1
	-2q_{0}(x,t) \right]\partial_y g(x,t \vert y,\tau).
\end{equation*}
We perform the integral over $x$ (using $\partial_y g(x,t \vert y,\tau)=-\partial_x g(x,t \vert y,\tau)$ and integration by parts) and get
\begin{equation}
	\int_{-\infty}^{\infty}dx\left[ 1-2q_{0}(x,t) \right]\partial_y g(x,t \vert y,\tau)=2g(y,2t-\tau \vert 0,0),
\end{equation}
and we substitute $q_1(y,\tau)$ from Eq.\eref{eq:q1 new} to yield
\begin{eqnarray*}
I_{1}-2I_{3}=8\int_{0}^{T}dt \int_0^t d\tau ~\tau \int_0^1 dr \\
\qquad ~ \int_{-\infty}^{\infty}dy\, g(y,\tau \vert 0,0) \erf{\frac{y \sqrt{r}}{2\sqrt{\tau (2-r)}}}\erf{\frac{y}{2\sqrt{\tau}}} g(y,2t-\tau\vert 0,0). 
\end{eqnarray*}
We integrate over $y$ by using the identity\footnote{Identity \eref{ab} appears in Ref.~\cite{Prudnikov}. One can also establish the validity of \eref{ab} using the same method as in the derivation of \eref{Ia_sin}. First, one notices that \eref{ab} is valid for $a=0$ or $b=0$. For general values of the parameters, it can be proved by showing that the derivatives of both sides with respect to $a$ are equal.} 
\begin{equation}
\label{ab}
\fl ~ ~ \int_{-\infty}^{\infty}dy~\exp[-\alpha y^2]~\erf{a~y}~\erf{b~y}=\frac{2}{\sqrt{\pi~\alpha}}\arctan\left[{\frac{a~b}{\sqrt{\alpha(a^2+b^2+\alpha)}}}\right]
\end{equation}
and we obtain 
\begin{equation}
\fl \qquad I_1-2I_3=\frac{8}{\pi^{3/2}} \int_0^T \frac{dt}{\sqrt{2t}}\int_0^t d\tau~\tau \int_0^1 dr\arctan \left[ \frac{(2t-\tau)\sqrt{r}}{2\sqrt{t(4t-tr-\tau)}}\right].
\end{equation}
The $T$ dependence can be extracted by defining a rescaled variable $s=\tau/t$ and integrating over $t$:
\begin{equation}
	I_{1}-2I_{3}=\frac{8~T^{5/2}}{5\sqrt{\pi}}\left[\frac{\sqrt{2}}{\pi}\int_0^1 ds~s \int_0^1 dr ~\arctan\left\{\frac{(2-s)\sqrt{r}}{2\sqrt{4-r-s}} \right\}\right].
\end{equation}
The last two integrals can be computed in \textit{Mathematica} to give
\begin{equation}
I_{1}-2I_{3}=\frac{8~T^{5/2}}{5\sqrt{\pi}}\left[-\frac{1}{6\sqrt{2}}+\frac{1}{\pi}\left\{ -2\sqrt{2}\arccos\left(\frac{5}{3\sqrt{3}} \right)+\frac{4}{3} \right\}\right].
\end{equation}

\subsection*{Computation of $I_2$}
Substituting $q_{1}(x,t)$ from Eq.\eref{eq:q1 final} into $I_2$ we obtain 
\begin{eqnarray}
\label{long}
 	I_{2}=4\int_{0}^{T}dt\int_{0}^{t}d\tau_{1}\int_{-\infty}^{\infty}dy_{1}\int_{0}^{t}d\tau_{2}\int_{-\infty}^{\infty}dy_{2}
~\erf{\frac{y_1}{2\sqrt{\tau_1}}}g(y_1,\tau_1\vert 0,0)\nonumber\\
\qquad\quad	\erf{\frac{y_2}{2\sqrt{\tau_2}}}g(y_2,\tau_2\vert 0,0)\int_{-\infty}^{\infty}dx ~ g(x,t\vert y_{1},
\tau_{1})~g(x,t\vert y_{2},
\tau_{2}).
\end{eqnarray}
Computing a Gaussian integral over $x$ we simplify Eq.\eref{long} to 
\begin{eqnarray}
I_{2}=4\int_{0}^{T}dt\int_{0}^{t}d\tau_{1}\int_{-\infty}^{\infty}dy_{1}\int_{0}^{t}d\tau_{2}\int_{-\infty}^{\infty}dy_{2}
~\erf{\frac{y_1}{2\sqrt{\tau_1}}}g(y_1,\tau_1\vert 0,0)\nonumber\\
\qquad\quad	\erf{\frac{y_2}{2\sqrt{\tau_2}}}g(y_2,\tau_2\vert 0,0)	g(y_{1}-y_{2}, 2t-\tau_{1}-\tau_{2}\vert
 0,0).
\end{eqnarray}

We extract the $T$ dependence by defining new variables
$\xi=y_{1}/2\sqrt{\tau_{1}}$, $\eta=y_{2}/2\sqrt{\tau_{2}}$, $m=\tau_{1}/t$ and
$n=\tau_{2}/t$: 
\begin{eqnarray}
\label{I1_long}
 I_{2}=T^{5/2}
	~\frac{4}{5\pi^{3/2}}\int_{0}^{1}dm\int_{0}^{1}dn\nonumber\\
		\qquad\qquad \int_{-\infty}^{\infty}d\xi\int_{-\infty}^{\infty}d\eta
	~\erf{\xi}e^{-\xi^{2}}\left[\frac{e^{
	-\frac{\left(\xi\sqrt{m}-\eta\sqrt{n}\right)^2}{2-m-n}
	}}{\sqrt{2-m-n}}\right]e^{-\eta^{2}}\erf{\eta}.
\end{eqnarray}
To integrate over $\xi$ and $\eta$ we use an integral representation of the error function 
\begin{equation}
	\erf{x}=\frac{2}{\sqrt{\pi}}\int_{0}^{1}dr~x\,\exp\left[-x^{2} r^{2}\right]
	\label{eq:int rep}
\end{equation}
and find
\begin{eqnarray}
\int_{-\infty}^{\infty}d\xi\int_{-\infty}^{\infty}d\eta~\erf{\xi}e^{-\xi^{2}}\left[\frac{e^{-\frac{\left(\xi\sqrt{m}-\eta\sqrt{n}\right)^2}{2-m-n}}}{\sqrt{2-m-n}}\right]e^{-\eta^{2}}\erf{\eta}\nonumber\\
\qquad\quad=\int_0^1dr\int_0^1ds\,\frac{2\,\sqrt{m~n}}{[ 2+ (2-m)r^2+(2-n)s^2+(2-m-n)r^2 s^2]^{3/2}}
\nonumber\\
\qquad\quad=\sqrt{2}~\arcsin\left[\sqrt{\frac{m~n}{(4-m)(4-n)}} \right].
\end{eqnarray}
This allows us to simplify Eq.\eref{I1_long} to 
\begin{equation}
	I_{2}=T^{5/2} \frac{4}{5\sqrt{\pi}}\left\{\frac{\sqrt{2}}{\pi}
	\int_{0}^{1}dm\int_{0}^{1}dn ~\arcsin\left[\sqrt{\frac{m~n}{(4-m)(4-n)}}
	\right]\right\}.
\end{equation}
The remaining double integral is computable, leading to the final result
\begin{equation}
	I_{2}=T^{5/2}\frac{4}{5\sqrt{\pi}}\left[
	 -\frac{5}{6}\sqrt{2} + \frac{1}{\pi} \left\{ 3\sqrt{2}\arccos\left(
       \frac{1}{3} \right)-\frac{4}{3} \right\}\right].
\end{equation}

\subsection*{Computation of $I_3$}
Using Eq.\eref{eq:partial p2} for $\partial_x p_2(x,t)$ and Eq.\eref{eq:sigma012} for $\sigma_0$ we re-write $I_3$ as
\begin{eqnarray}
\label{I3:long} 
 I_3=\int_{0}^{T}dt\int_{t}^{T}dt_{1}\int_{t}^{T}dt_{2}\int_{-\infty}^{\infty}dx~\erfc{\frac{x}{2\sqrt{t}}}\erfc{-\frac{x}{2\sqrt{t}}}\nonumber\\
	 \qquad\qquad\qquad  \qquad\qquad  \qquad\qquad g(0,2t_{1}\vert x,t)~g(0,2t_{2}\vert x,t).
\end{eqnarray}

To extract the $T$ dependence we use new variables 
$\xi=x/2\sqrt{t}$, $r=2t_{1}/t$, $s=2t_{2}/t$ and $\tau=t/T$ and transform Eq.\eref{I3:long} to 
\begin{eqnarray}
I_{3}=T^{5/2}\left[ \int_{0}^{1}d\tau
	\left(\frac{\tau^{3/2}}{8\pi}\right)\int_{2}^{2/\tau}dr\int_{2}^{2/\tau}ds
	~\frac{1}{\sqrt{(r-1)(s-1)}}\right.\nonumber\\
	 \qquad \qquad
	 \qquad\left.\int_{-\infty}^{\infty}d\xi~\erfc{\xi}~\erfc{-\xi}e^{-\left(
	\frac{1}{r-1}+\frac{1}{s-1}\right)\xi^2}\right].
\end{eqnarray}
The integration over $\xi$ can be performed using the following identity~\footnote{Identity \eref{EEA} appears in \cite{Prudnikov}. Alternatively, one can establish it by using the integral representation \eref{eq:int rep}, computing the resulting Gaussian integral and evaluating the remaining algebraic integral.}
\begin{equation}
\label{EEA}
	\int_{-\infty}^{\infty}d\xi ~\erfc{\xi} ~\erfc{-\xi}
	\exp\left[-\alpha~\xi^2\right]=\frac{2}{\sqrt{\pi\alpha}}~\arcsec{
	1+\alpha },
\end{equation}
which is valid for all for $\alpha>0$. This gives 
\begin{eqnarray}
\fl \qquad	I_{3}=\frac{T^{5/2}}{\sqrt{\pi}}\left[ \int_{0}^{1}d\tau
\left(\frac{\tau^{3/2}}{4\pi}\right)\int_{1}^{\frac{2}{\tau}-1}du\int_{1}^{\frac{2}{\tau}-1}dv
	~\frac{\arcsec{1+\frac{1}{u}+\frac{1}{v}}}{\sqrt{u+v}}\right],
\end{eqnarray}
where we have defined $u=r-1$ and $v=s-1$. We computed this last set of integrals in \textit{Mathematica} and we got
\begin{equation}
	I_{3}=T^{5/2}~\frac{4}{15\sqrt{\pi}}\left[ -4+3\sqrt{2} \right].
\end{equation}

Combining the results for the three integrals $I_{1}$, $I_2$ and $I_{3}$ in the
expression \eref{eq:A4 sum} for the fourth cumulant yields the answer reported
in Eq.\eref{eq:A4 final}.

\section{Fluctuating hydrodynamics \label{app:fluc hyd}}
In general, the fluctuation of an interface with an underlying microscopic dynamics
satisfying detailed balance, is  modeled by a continuum description in terms of the Edwards-Wilkinson equation
\cite{Stanley,Halpin1995,book}. This is a linear stochastic partial differential equation
describing the time evolution of a continuous height profile $h(x,t)$, as
\begin{equation}
	\partial_{t}h(x,t)=\partial_{xx}h(x,t)+\eta(x,t).
\end{equation}
The $\eta(x,t)$ is a white noise with covariance
\begin{equation}
	\langle \eta(x,t)\eta(x^{\prime},t^{\prime})\rangle= \Gamma
	~\delta(x-x^{\prime})~\delta(t-t^{\prime}),
\end{equation}
with $\Gamma$ being a parameter.
In the framework of the SSEP, this corresponds to the equation for the density field,
\begin{equation}
	\partial_t\rho(x,t)=\partial_{x}\left[\partial_{x}\rho(x,t)+ \eta(x,t)\right],
	\label{eq:EW}
\end{equation}
where we have used the relation (\ref{eq:h tau relation})  between the height variable and the density 
in the continuous limit. The above  dynamics is mass conserving: $\int dx\,\rho(x,t)$ remains constant.

The scaling of the fluctuations of $d_{T}$ and $A_{T}$ in the Ising quadrant can be argued within
the Edwards-Wilkinson description. The interface fluctuations are
characterized by the scaling exponents $\zeta$ and $z$ which are defined by $W_{T}\sim
T^{\zeta/z}$ and $\ell_{T} \sim T^{1/z}$,  where $W_{T}$ is the width of the
interface and $\ell_{T}$ is the height-height correlation length at time $T$. For the
$(1+1)$ dimensional Edwards-Wilkinson interface, $\zeta=1/2$ and $z=2$ (the diffusive growth) \cite{Stanley}. 
 In our case of the Ising quadrant, as the interface is
unbounded, the correlation length $\ell_{T}$ sets the characteristic length
scale. Therefore, because of the diffusive growth $\langle d_{T} \rangle \sim \sqrt{T}$ and $\langle A_{T}
\rangle \sim T$. On the other hand,
the variance of the diagonal height is given by $\langle d_{T}^{2} \rangle_{c}\sim W_{T}^{2} \sim
\sqrt{T}$. To find results for the variance of $A_{T}$ we note that the interface can
be considered as made of segments of length $\ell_{T}$ such that there is almost no correlation between the height fluctuations in two segments. There are $\sim \sqrt{T}/\ell_{T}$ such
segments and we conclude that 
\begin{equation*}
  \langle A_{T}^{2} \rangle_{c}\sim \left( \frac{\sqrt{T}}{\ell_{T}} \right)
  \left( W ~\ell_{T} \right)^2 \sim
  T^{\frac{1}{2}+\frac{1}{z}+\frac{2\zeta}{z}}.
\end{equation*}
Substituting the values of $\zeta$ and $z$ yields the correct $T$ dependence $\langle
A_{T}^{2} \rangle_{c}$ as in Eq.\eref{eq:A2_first}. This simple argument can be generalized to higher
dimensions as well, as done for a crystal growth problem in \cite{Olejarz2013}.

One can go further  and compute the exact expressions of the average and the variance of $d_{T}$ and $A_{T}$ from this Edwards-Wilkinson description Eq.~\eref{eq:EW}. A simple analysis \cite{Krapivsky2012} yields the
correct expression for $\langle d_T \rangle$, $\langle d_{T}^{2} \rangle_{c}$
and $\langle A_{T} \rangle$. However, the variance $\langle A_{T}^{2}\rangle_{c}$ calculated using Eq.~\eref{eq:EW} diverges at any non-zero time $T$. This is related to the fact that the noise
amplitude $\Gamma$ is non-vanishing even far from the origin. Hence, a blindfolded application of the Edwards-Wilkinson equation is inadequate for the integral properties of the interface  and we must  use the correct noise
amplitude. This is achieved by taking  the noise amplitude to be  $\sigma(\rho)$ in
Eq.~\eref{eq:sigma} which vanishes when the density $\rho$ is zero or one (far from the origin in our setting). 
This qualitatively explains the virtues of the Langevin equation discussed in \sref{sec:mft}. We have shown that MFT allows us to calculate perturbatively  the cumulants. However, the analysis is tedious and involved. If we are interested only in calculating  the variance of $A_{T}$, it can be  obtained in a rather simpler setting by assuming small fluctuations and linearizing Eq.\eref{eq:fh} around the hydrodynamic solution. We now explain briefly how this can be done.
 
Let us  write $\rho(x,t)=\rho_{0}(x,t)+u(x,t)$, where the deterministic part 
$\rho_{0}(x,t)$ is the solution of the diffusion equation $\partial_t\rho_{0}=\partial_{xx}\rho_{0}$, i.e., 
$\rho_{0}(x,t)=2^{-1}\erfc{x/\sqrt{4t}}$ in our case of the step initial condition. The fluctuating field $u(x,t)$ satisfies
\begin{equation}
\label{uxt}
	\partial_{t}u=\partial_{xx}u+\partial_{x}\left[
	\sqrt{\sigma(\rho_{0})}~\eta(x,t)\right] \, ,
\end{equation}
with the initial condition  $u(x,0)=0$. We have assumed that $u(x,t)$ is small.
Note that, with this linearisation, the noise amplitude does not depend on the fluctuating field
$u(x,t)$. Solving Eq.\eref{uxt} we obtain 
\begin{equation}
	u(x,t)=\int_{0}^{t}d\tau \int_{-\infty}^{\infty}dy~\sqrt{
	\sigma\left[\rho_{0}(y,\tau)\right]}~\eta(y,\tau)~\partial_{x}g(x,t\vert y,\tau),
	\label{eq:u}
\end{equation}
where $g(x,t\vert y,\tau)$ is the diffusion propagator (\ref{eq:g}). Note that
 $\langle u(x,t)\rangle =0$ because  $\langle
\eta(x,t)\rangle=0$. Using this together with Eq.\eref{eq:AT definition} one can check that
\begin{equation}
	\langle A_{T} \rangle =\int_{-\infty}^{\infty}dx
	\left[\rho_{0}(x,T)-\rho_{0}(x,0) \right] x = T .
\end{equation}
Similarly, the variance is given by 
\begin{equation}
	\langle
	A_{T}^{2}\rangle_{c}=\int_{-\infty}^{\infty}dx\int_{-\infty}^{\infty}dy
	~x y~\langle u(x,T)~u(y,T) \rangle.
\end{equation}
Using the solution for $u(x,t)$ from Eq.\eref{eq:u} the above expression simplifies to
\begin{equation}
	\langle A_{T}^{2}\rangle_{c}=\int_{0}^{T}dt \int_{-\infty}^{\infty}dx
	~\sigma\left[\rho_{0}(x,t)\right],
\end{equation}
which is identical to  Eq.~\eref{eq:A2} found previously by using the MFT.

We emphasize that the assumption of small fluctuations around the hydrodynamic profile does not give the correct results for higher cumulants. Thus one has to resort to the more detailed perturbative analysis as described in \sref{sec:mft}.

\section{Half-area \label{app:half-area}}

In this appendix, we consider another observable $H_{T}$ which is the
sum of the position of all the particles on the positive half-line at time
$T$. In terms of the occupation  variables,  this corresponds to
\begin{equation}
	H_{T}=\sum_{x=0}^{\infty}~ n_{x}(T)~ x.
\end{equation}
Then, as long as the positive half-line is empty in the initial configuration,
the $H_{T}$ remains finite at a finite $T$.
Going back to the $45$ degree-rotated Ising quadrant picture (see \Fref{fig:def-new}), this
quantity $H_{T}$ is the area of the molten region below the diagonal $u=v$ line.
On an average,   $H_{T}$  is  the half of the total area $A_{T}$.

There are two  advantages of studying  $H_{T}$ rather than $A_{T}$. First, we can
consider more than one initial profile as long as the positive half-line is
empty;  second, $H_{T}$ can be calculated for  non-interacting particles. 

In several one-dimensional systems away from their stationary state there is
a non-trivial dependence on the initial state, even at large times. This effect
has recently been observed in the statistics of $Q_{T}$ in the SSEP
\cite{DG09} and in the large deviation of the position of a tagged particle in
the single-file diffusion \cite{Krapivsky2014}. The fluctuations of
$H_{T}$ in our problem also exhibits such sensitivity to the
initial state. We consider two specific initial profiles: (a) annealed, where the
distribution of the particles on the negative half-line is fluctuating, and (b) quenched,
where the initial configuration is fixed. In both cases the positive line is
empty and the average density on the negative line is $\rho$. In a detailed calculation using macroscopic fluctuation theory we
find that in both cases the average value of $H_{T}$ is same and equal to
$\rho~T/2$; however, the difference appears in the variance,
\begin{equation}
	\langle H_{T}^{2} \rangle = \cases{\frac{2}{\sqrt{\pi}}\left[
	\frac{3-\sqrt{2}}{3}\rho -\frac{2-\sqrt{2}}{3}\rho^{2} \right]\times
	T^{3/2} & for quenched,\\ \frac{2}{\sqrt{\pi}}\left[
	\frac{2}{3}\rho -\frac{1}{3}\rho^{2} \right]\times
	T^{3/2} & for annealed,}
	\label{eq:H var}
\end{equation}
(the calculation is similar to the one used for the variance of integrated current in \cite{Krapivsky2012}, or  the variance of tagged particle position in single-file diffusion \cite{Krapivsky2014}). Note, that for $\rho=1$, where due to exclusion, both annealed and quenched
cases have the same
initial configuration, the two expressions for the variance match.
At this density,  $A_{T}$ is well defined and
 an interesting feature emerges;  the average $\langle A_{T} \rangle$ is twice that of average $\langle
H_{T} \rangle$, as expected. However, the variance $\langle A_{T}^{2}
\rangle_{c}=2\sqrt{2} ~\langle H_{T}^{2} \rangle_{c} $. Had the fluctuations of
the interface on the two
sides been independent, the relation would have $\langle A_{T}^{2} \rangle_{c}=2 \langle
H_{T}^{2} \rangle_{c} $. On the other-hand for a fully correlated case the pre-factor
would have been $4$. The intermediate value $2\sqrt{2}$ as found, indicates a
non-trivial correlation between the fluctuations
of the interface on the two sides of the diagonal. 

As mentioned earlier, the quantity $H_{T}$ is well defined even for
non-interacting particles.
With time, each particle performs an unbiased random walk on the lattice with symmetric jump rates $1$,
independent of the others.
An advantage of the non-interacting particles is that the analysis is simpler and
the cumulant generating function can be determined for both annealed and
quenched initial conditions.

To illustrate the derivation, we consider the simplest case of a step initial profile:
all sites on the negative half-line including the zeroth site are occupied, \textit{i.e},  $\rho=1$. Let $y_j(T)$ is the position of a particle at time $T$ which started at site $-j$ at time $t=0$. By definition, the half-area $H_T=\sum_{j=0}^{\infty}y_j(T)~\Theta[y_j(T)]$, where  $\Theta(x)$ is the Heaviside step function. As the particles are independent of each other, the generating function of $H_{T}$ can be simplified as
\begin{equation}
  \langle e^{\lambda H_{T}} \rangle=\langle e^{\lambda \sum_{j=0}^{\infty}y_j(T)~\Theta[y_j(T)]} \rangle=\prod_{j=0}^{\infty} \langle e^{\lambda
    y_{j}(T)~\Theta[y_j(T)]} \rangle,
    \label{eq:ni gen 1}
\end{equation}
where the angular brackets denote average over the history.

The $y_j(T)$ is a random variable depending on the history of the $j$th particle. It is easy to show that
\begin{eqnarray}
  \langle e^{\lambda y_{j}(T)~\Theta[y_j(T)]} \rangle &=& \sum_{y=1}^{\infty}e^{\lambda y}\mathcal{P}_{T}(j\vert y)+1-\sum_{y=1}^{\infty}\mathcal{P}_{T}(j\vert y)\nonumber\\
  &=& 1+\sum_{y=1}^{\infty}\left( e^{\lambda y}-
  1\right)\mathcal{P}_{T}(j\vert y),
\end{eqnarray}
where $\mathcal{P}(j\vert y)$ is the probability that a particle started at
$-j$ ends up at $y$ at time $T$. For a random walker
\begin{equation}
  \mathcal{P}_{T}(j\vert y)=\frac{1}{\sqrt{4\pi T}}\exp\left[ -
  \frac{(j+y)^{2}}{4T} \right].
\end{equation}
Substituting this result in Eq.\eref{eq:ni gen 1} yields
\begin{equation}
  \langle e^{\lambda H_{T}} \rangle =\prod_{j=0}^{\infty} \left[1+\sum_{y=1}^{\infty}\left( e^{\lambda y}-
  1\right)\frac{e^{-\frac{(j+y)^{2}}{4T} }}{\sqrt{4\pi T}} \right].
 \end{equation}

One can extend the analysis for initial state with any general density $\rho>0$. The result depends on the type of the initial states considered: quenched or annealed.
In the continuum limit, the cumulant generating function yields
\begin{equation}
\kappa_{T}(\lambda) =\ln\langle e^{\lambda H_T} \rangle= \cases{
\rho \int_{0}^{\infty}dx\ln\left[ 1+ \Phi(\lambda, x, T)\right] & quenched,\\
\rho\int_{0}^{\infty}dx~\Phi(\lambda, x, T)  & annealed.}  
\end{equation}
where we have used the shorthand notation
\begin{equation*}
\Phi(\lambda, x, T) = \int_{0}^{\infty}dy \left( e^{\lambda y}-1\right)\frac{e^{-\frac{(x+y)^{2}}{4T}}}{\sqrt{4\pi T}}.
\end{equation*}
In the annealed case we considered an equilibrium initial state where a site on the negative half-line is populated following a Poisson distribution of average density $\rho$. The results for $\kappa_T(\lambda)$ is different for other fluctuating initial state.

\section*{References}

\bibliographystyle{unsrt}
\bibliography{reference}

\end{document}